# MR-Contrast-Aware Image-to-Image Translations with Generative Adversarial Networks

Jonas Denck[123], Jens Guehring[3], Andreas Maier[1], Eva Rothgang[2]

[1] Pattern Recognition Lab, Department of Computer Science, Friedrich-Alexander Universität Erlangen-Nürnberg, Erlangen, Germany

[2] Department of Industrial Engineering and Health, Technical University of Applied Sciences Amberg-Weiden, Weiden, Germany

[3] Siemens Healthineers, Erlangen, Germany

## Abstract

**Purpose**
A Magnetic Resonance Imaging (MRI) exam typically consists of several sequences that yield different image contrasts. Each sequence is parameterized through multiple acquisition parameters that influence image contrast, signal-to-noise ratio, acquisition time, and/or resolution. Depending on the clinical indication, different contrasts are required by the radiologist to make a diagnosis. As MR sequence acquisition is time consuming and acquired images may be corrupted due to motion, a method to synthesize MR images with adjustable contrast properties is required.

**Methods**
Therefore, we trained an image-to-image generative adversarial network conditioned on the MR acquisition parameters repetition time and echo time. Our approach is motivated by style transfer networks, whereas the "style" for an image is explicitly given in our case, as it is determined by the MR acquisition parameters our network is conditioned on.

**Results**
This enables us to synthesize MR images with adjustable image contrast. We evaluated our approach on the fastMRI dataset, a large set of publicly available MR knee images, and show that our method outperforms a benchmark pix2pix approach in the translation of non-fat-saturated MR images to fat-saturated images. Our approach yields a peak signal-to-noise ratio and structural similarity of 24.48 and 0.66, surpassing the pix2pix benchmark model significantly.

**Conclusion**
Our model is the first that enables fine-tuned contrast synthesis, which can be used to synthesize missing MR contrasts or as a data augmentation technique for AI training in MRI.

## Introduction

A Magnetic Resonance Imaging (MRI) exam typically consists of multiple sequences that yield different image tissue contrasts required for a complete and reliable diagnosis. However, the sets of sequences that are obtained vary considerably across clinical protocols, scanners, and sites. Clinical guidelines, the MR system (vendor, model, software version, field strength), internal guidelines (e.g., slot time), and radiologists' preferences determine the set of selected sequences (i.e., the MRI protocol) for a specific clinical question at a particular site. Moreover, each sequence is parameterized through multiple acquisition parameters (pulse sequence parameters) that affect image contrast, image resolution, signal-to-noise ratio, and/or acquisition time. The acquisition parameters (and sequences) can be proprietary or generic across vendors. Consequently, sequence parameterization and selection vary significantly across different radiology sites and within exams and scanners of the same site [1].

As sequence acquisition is time consuming and acquisition time is expensive, current research strives to increase MRI value [2] by, e.g., the development of optimized scan protocols [3] or the reduction of scan time by leveraging artificial intelligence (AI) [4]. AI plays a vital role in reducing scan time by either accelerating image acquisition and reconstruction [5] or synthesizing missing or corrupted image contrasts from existing ones [6–8]. The latter is typically done using (multi-)image-to-image neural networks, e.g., to synthesize T2-weighted brain images from T1-weighted images and vice versa [6, 9]. This can offer great clinical value as corrupted images due to motion or other artifacts can be replaced, or claustrophobic patients that prematurely had to end the MRI scan may avoid re-scans. Current approaches synthesize a single MR contrast with fixed or not further specified acquisition parameters from one or multiple existing MR contrasts. These approaches have been trained and tested on different publicly available MR datasets. However, as sequence parameterizations vary in the clinical practice, these approaches are only applicable to a small share of used sequences. To truly increase the clinical value, these approaches must synthesize MR contrasts beyond fixed sequence parameterizations.

Therefore, we have developed and trained an image-to-image generative adversarial network (GAN) that synthesizes MR images with adjustable image contrast. The acquisition parameters echo time (TE) and repetition time (TR) that influence image contrast are incorporated into the network's training. Moreover, we can translate non-fat-saturated MR images into fat-saturated MR images with the acquisition parameters TE and TR as additional inputs, which increases the overall reconstruction performance. We provide a thorough visual and quantitative evaluation of our approach and benchmark it with the commonly used pix2pix framework.

## Material and Methods

### Generative Adversarial Network

In this section, we want to give a short overview of generative adversarial networks and important adaptions for medical image-to-image synthesis.

GANs are a special type of artificial neural network where two networks (generator and discriminator) are trained adversarially. For an image generation task, the generator is focused on image generation, and the discriminator learns to discriminate between real and generated (fake) images.

An important extension of GAN capabilities is pix2pix, which serves as a general-purpose solution to image-to-image translation problems. The pix2pix approach is used to learn the mapping from an input image to a paired output image under an L1 reconstruction and an adversarial loss [10].

Since paired training data are not available for many image-to-image translation tasks, cycle consistency loss was proposed (CycleGAN) [11] that allows the translation of an image from a source domain to a target domain without any paired training examples. The original formulation includes two generative models, G and F, where G translates an image from domain A into domain B. F translates an image from domain B into domain A. The discriminators $D_A$ and $D_B$ learn to distinguish between real and fake images from their domains. The cycle consistency loss using the L1 distance for image g and t from domain A and B, respectively, is then defined as:

$$L_{Cycle} = E[\|G(F(g)) - g\|_1] + E[\|F(G(t)) - t\|_1] \tag{1}$$

Both pix2pix and CycleGAN are the basis for many image-to-image translation tasks in the medical imaging domain [7–9].

### Image-to-Image Generative Adversarial Network

Image-to-image GANs, e.g., pix2pix, mainly focus on the pixel-to-pixel image synthesis. This generally leads to an overall low pixel intensity error, but it can also lead to neglecting contrast characteristics of an image, as minimizing a pixel-intensity-based reconstruction loss does not necessarily produce the correct image contrast.

Therefore, we incorporate a loss term that penalizes contrast differences through the conditioning loss of an auxiliary classifier. The auxiliary classifier is pretrained to determine the acquisition parameters that determine the contrast.

In order to produce sharp and realistic images, we use the non-saturating adversarial loss with R1 regularization using $\gamma = 1$ [12].

Additionally, we use a L1 reconstruction loss to enforce pixel-wise similarity between target ground truth and reconstructed image. However, an important characteristic for MRI is that MR images present significant intensity variation across patients and scanners, and MR image intensity standardization is an ongoing research topic [13]. Hence, a reconstruction loss that is not fully focused on pixel intensity similarities, but perceptually motivated, is anticipated to improve the performance of the image-to-image GAN. Therefore, we also experimented with a weighted reconstruction loss of L1 and multi-scale structural similarity (MS-SSIM) index, originally proposed by [14]. MS-SSIM preserves the contrast in high-frequency regions, but is not sensitive to uniform biases, i.e., can produce brightness shifts, while L1 preserves pixel intensities [14]. Given two images $\mathbf{x}$ and $\mathbf{x}'$, the MS-SSIM loss is defined as:

$$\mathcal{L}^{MS\text{-}SSIM}(\mathbf{x}, \mathbf{x}') = 1 - MS\text{-}SSIM(\mathbf{x}, \mathbf{x}') \tag{2}$$

where MS-SSIM is the multiscale version of SSIM (7) that is defined in the Section "Evaluation Metrics". The weighted reconstruction loss is then given as:

$$\mathcal{L}^{Recon}(x, x') = \omega \cdot \mathcal{L}^{MS\text{-}SSIM}(x, x') + (1 - \omega) \cdot \mathcal{L}^{Recon}(x, x') \tag{3}$$

with $\omega = 0.84$. The value for α is set to balance the contribution of the two loss terms and proposed by [14].

## Adaptive Instance Normalization

We use adaptive instance normalization [15] to inject the input and output labels (TE, TR, fat saturation) into the generator. The AdaIN operation is defined as:

$$AdaIN(x, y) = \alpha(y) \cdot \left(\frac{x - \mu(x)}{\sigma(x)}\right) + \beta(y) \tag{4}$$

Each feature map $x$ is normalized separately with its mean $\mu(x)$ and standard deviation $\sigma(x)$, then scaled and biased through learned affine transformations $\alpha, \beta$, given the label set $y$.

AdaIN has already been applied successfully for style transfer applications by injecting style encodings into the generator [15]. The style encoding is typically unknown for style transfer and is learned implicitly from an image [16]. In our case, the "style" can be understood as image tissue contrast, which is known as it is defined through the acquisition parameters. Thus, in MR imaging, the "style" encoding is given explicitly and does not have to be learned.

## Auxiliary Classifier

We deviate from the conventional Auxiliary Classifier GAN (ACGAN) network architecture [17], which uses a classification layer in the discriminator to learn the conditions by employing a separate auxiliary classifier (AC) that is only trained on the conditions. This allows us to pretrain the AC and breaks down the training complexity as the AC performance can be tuned separately from the GAN. Only a well-trained auxiliary classifier can provide useful guidance for the generative model's training, crucial for good image reconstruction accuracy.

## Network and Training Details

The training procedure and network architecture are depicted in Fig. 1. The generator consists of a U-Net structure with residual blocks. It injects the source image acquisition parameters into the network's encoder and the target image's acquisition parameters (i.e., target contrast) into the network's decoder via adaptive instance normalization. Through the "style" injection in the encoder, the generator is anticipated to learn an image representation independent of the acquisition parameters, thus image contrast. The target acquisition parameters are then injected into the decoder to reconstruct the target contrast properly.

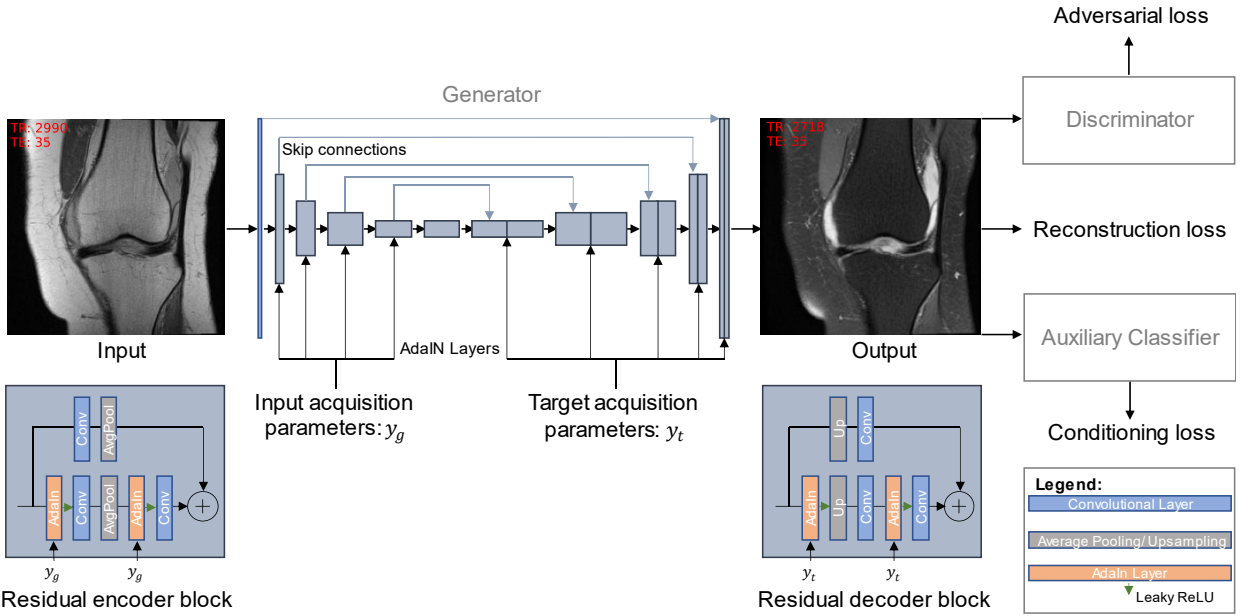

**Fig. 1** GAN network architecture. The generator is based on a U-Net architecture consisting of residual blocks with adaptive instance normalization. The input labels ($y_g$) are injected into the encoder part of the generator and the output target labels ($y_t$) are injected into the decoder part of the generator. The generator is trained on the reconstruction, adversarial, and conditioning loss.

The discriminator consists of six residual blocks and a single output to discriminate between real and synthetic images.

We use the recently published EfficientNet-B3 [18] architecture for the AC and train the network on the determination of TE and TR value and whether fat saturation was used. EfficientNet uses compound scaling and achieves higher accuracy and better efficiency over existing CNNs on the ImageNet and other benchmark datasets [18]. The mean squared error loss is used for TE and TR, whose values are both scaled to values between 0 and 1, and binary cross-entropy for determining the use of fat saturation.

The AC is pretrained for 200K iterations with a batch size of 64, using Adam optimizer with a learning rate of $10^{-4}$, $\beta_1 = 0$, and $\beta_2 = 0.99$.

We use a batch size of eight for training the GAN and train the model for 200K iterations. We also use the Adam optimizer with $\beta_1 = 0$ and $\beta_2 = 0.99$. The learning rates for the generator and discriminator are set to $10^{-4}$. For the generator, exponential moving averages over its network parameters are employed [19, 16].

The images are scaled to intensities of [-1, 1] and resized to a resolution of 256×256 pixels using bilinear interpolation to obtain an identical image resolution within the dataset. Random image shifting and zooming are applied as data augmentation during the training of the AC and GAN.

## Data

We used the fastMRI dataset [4] for our training and evaluation. It contains DICOM data from 10,000 clinical knee MRI studies, each comprising a set of multiple sequence parameterizations. We applied several data filters based on the DICOM header information to get a dataset with a comparable image impression, a dense and homogenous acquisition parameter distribution (TE, TR), and high variance in anatomy. We wanted the image impression and contrast within our training set to depend on the acquisition parameters TE and TR.

Therefore, other parameters affecting the image impression were removed, such as field strength and manufacturer, by selecting the most common parameter value within the fastMRI dataset (1.5T field strength and scanners from Siemens Healthcare, Erlangen, Germany). The MR images (fast-spin-echo sequences) from our filtered dataset were acquired on five different Siemens Healthcare scanners (MAGNETOM Aera, MAGNETOM Avanto, MAGNETOM Espree, MAGNETOM Sonata, MAGNETOM Symphony). We excluded MR images with a TR over 5000 ms and set the upper limit of TE to 50 ms to create a dataset with a dense representation of the conditions. Fig. 2 shows the distribution of TE and TR values in the training dataset. Moreover, we took the seven central slices from each volume to discard peripheral slices.

Image pairs were then defined as images with identical DICOM attributes patient ID (0010,0020), study instance UID (0020,000D), image orientation (0020,0037), slice location (0020,1041), and slice thickness (0018,0050).

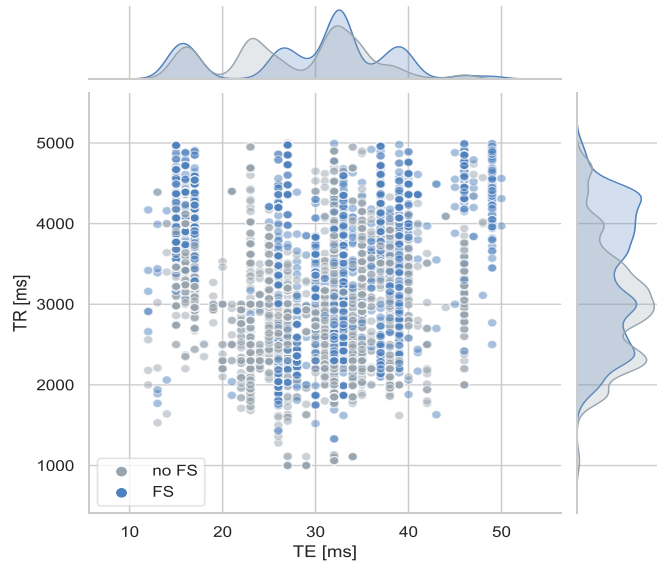

**Fig. 2** Distribution of the acquisition parameters TE and TR in the training dataset, color coded by fat saturation (FS).

# Results

## Data

The training dataset consists of 237,883 MR images, with 101,328 paired images and 136,555 unpaired images, from 4,815 different studies and 16,731 image series. The dataset contains 123,833 fat-saturated MR images, and the remaining images were acquired without fat saturation.

The validation and the test dataset consist of paired images from 100 disjunct, randomly selected patient IDs (3,242 and 3,438 images in total, respectively).

## Evaluation Metrics

For the evaluation of our experiments, which are described in the following, we used the normalized mean squared error (NMSE), peak signal-to-noise ratio (PSNR), and structural similarity index (SSIM). Given two images $\mathbf{x}$ and $\mathbf{x}'$, they are defined as:

$$\text{NMSE}(\mathbf{x}, \mathbf{x}') = \frac{\|\mathbf{x} - \mathbf{x}'\|_2^2}{\|\mathbf{x}\|_2^2} \tag{5}$$

$$\text{PSNR}(\mathbf{x}, \mathbf{x}') = 10 \cdot log_{10}\left(\frac{MAX_{range}^2(\mathbf{x}, \mathbf{x}')}{\frac{1}{n}\|\mathbf{x} - \mathbf{x}'\|_2^2}\right) \tag{6}$$

$$SSIM(\mathbf{x}, \mathbf{x}') = \frac{(2 \cdot \mu_x \mu_{x'} + c_1) \cdot (2 \cdot \sigma_{xx'} + c_2)}{(\mu_x^2 + \mu_{x'}^2 + c_1) \cdot (\sigma_x^2 + \sigma_{x'}^2 + c_2)} \tag{7}$$

where $n$ is the number of pixels in one image, MAX is the maximum pixel value range of two images, $\mu_x$ and $\mu_{x'}$ denote the mean values of original and translated images. $\sigma_x$ and $\sigma_{x'}$ denote the standard deviation of original and translated images, and $\sigma_{xx'}$ is the covariance of both images. The variables $c_{1,2}$ are added to stabilize the division with a weak denominator.

Lower values for NMSE and higher values for PSNR and SSIM demonstrate better image synthesis.

## Experiments

We have conducted multiple experiments to evaluate our approach thoroughly. First, we assess the AC's performance to determine the acquisition parameters TE and TR from the MR image, as its performance is crucial for the performance evaluation and the guidance for the generator to synthesize MR images. Then, we evaluate our GAN on the task of synthesizing fat-saturated (FS) images from non-fat-saturated images and benchmark it with the pix2pix approach. Moreover, we demonstrate the capability of our approach to synthesize MR images with adaptable image contrast.

We evaluated several image-to-image GANs. Each model is based on the prior model with the additional changes as described in the following:

- Model 1: pix2pix – the training data consist of one-directional image pairs, i.e., non-FS (non-fat-saturated) to FS image pairs, under a L1-reconstruction and non-saturating adversarial loss with R1 regularization.
- Model 2: target labels are injected into the decoder with AdaIN layers.
- Model 3: source labels are additionally injected into the encoder with AdaIN layers.
- Model 4: the L1 reconstruction loss is adapted based on Formula 3.
- Model 5: image pairs with non-fat-saturated target images are added to the training target data, enabling non-FS to FS and non-FS translations.
- Model 6: unpaired non-FS image data are added to the training dataset and trained under the cycle consistency and conditioning loss (with weight $\lambda_c = 10$). The network is trained on unpaired data according to its share in the dataset. $\lambda_c$ is a hyperparameter and is set heuristically such that the conditioning validation error is similar to the error of the auxiliary classifier (i.e., is not over- or underfitting w.r.t. the conditioning).

Please note the following notations that are used in the remainder of the evaluation: $g$ denotes the input ground truth image with the labels $y_g$, $t$ represents the target image with its labels $y_t$. The generator $G$ then translates the input image to the target contrast, denoted by $G(g, y_g, y_t)$.

The performance results on the test set are reported in Tables 3 and 4. A good performance of the auxiliary classifier is crucial for proper guidance for the generator (model 5) during training and the evaluation of all GAN models. Although the AC is trained under MSE loss, we report the mean absolute error as it is more meaningful for the performance evaluation (Table 3). The auxiliary classifier predicts the acquisition parameters correctly with a low overall error. TE has a stronger impact on image contrast than TR (compare Fig 4.), which might indicate why the MAE of the AC for TE is (proportionally) lower than for TR. Moreover, the fat-saturation distinction works perfectly (accuracy: 100%).

**Table 3** Evaluation of the auxiliary classifier. The reported values denote the mean and its standard deviation on the test set

| Model | TE [ms] | TR [ms] | FS [%] |
| --- | --- | --- | --- |
| AC (non-FS) | 1.8 ± 2.1 | 247 ± 262 | 100 |
| AC (FS) | 1.8 ± 2.2 | 290 ± 341 | 100 |
| 1 | 3.0 ± 3.4 | 447 ± 419 | <u>100</u> |
| 2 | 1.4 ± 1.7 | 354 ± 357 | <u>100</u> |
| 3 | 1.3 ± 1.7 | 298 ± 288 | <u>100</u> |
| 4 | <u>1.1 ± 1.5</u> | <u>261 ± 277</u> | <u>100</u> |
| 5 | 1.3 ± 2.5 | <u>260 ± 270</u> | <u>100</u> |
| 6 | 2.2 ± 2.0 | 313 ± 293 | <u>100</u> |

- MAE is reported for TE and TR and accuracy for FS.
- Statistically significant (p ≤ 0.05) best values for the models are underlined.
- Note: Since for model 1-3 only FS images can be synthesized, the conditioning evaluation for all models

We also evaluated the models' conditioning error on the test set (Table 3) to assess how well the target contrasts are synthesized. Therefore, we translated the input test images to the contrast settings of the paired target image, determined the acquisition parameters with the AC, and computed the error.

The benchmark pix2pix model (model 1) shows a high conditioning error. It is not guided by the input or target acquisition parameters and can only translate an image to the expected contrast. Incorporating the target acquisition parameters (model 2) and the input acquisition parameters (model 3) reduces the MAE on TE and TR determination, demonstrating a better generation of the target MR image contrast. Incorporating more training data (models 4 and 5) further enhances the contrast synthesis.

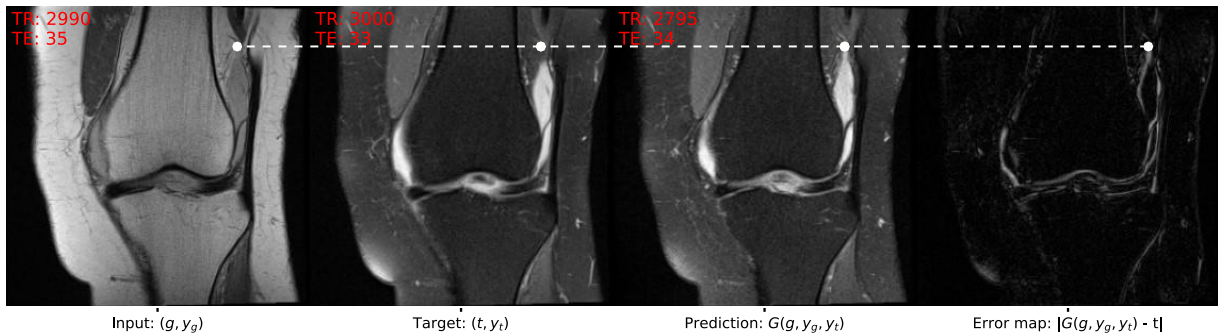

Input: $(g, y_g)$    Target: $(t, y_t)$    Prediction: $G(g, y_g, y_t)$    Error map: $|G(g, y_g, y_t) - t|$

**Fig. 3** Image-to-image translation example with our proposed approach (model 5). The figure shows a paired example of a real non-fat-saturated and fat-saturated MR knee image with the reconstructed MR images given the input image with corresponding input and output labels, $G(g, y_g, y_t)$, and the corresponding absolute error map. The images are annotated with the true acquisition parameters (TE, TR) for the input and target, and with the acquisition parameters determined by the AC for the reconstructed image $G(g, y_g, y_t)$. The white line indicates an example of imperfect image registration of the image pair of input and target image due to through-plane motion, reducing the dataset quality (see Discussions).

The reconstruction performance is reported in Table 4. Incorporating input and target acquisition parameters significantly increases the reconstruction performance (model 2 and 3) w.r.t. the benchmark model. Including more training data (model 4) further enhances the performance and extends the network's capabilities: it enables the model to interpolate the MR contrast within non-fat-saturated images (Fig. 4) while also allowing translation of non-fat-saturated into fat-saturated images. Introducing unpaired training data and cyclic reconstruction (model 5 and 6) allows the training on a larger set of input-target acquisition parameter sets (with randomly selected target acquisition parameters) that increases the model's generalization ability.

The reconstructed images in Fig. 3-5 were generated using model 6, as contrast changes can be recognized best with this model. The reconstruction of a fat-saturated image from its non-fat-saturated image pair is presented in Fig. 3. A novel functionality of our approach is the contrast interpolation capabilities shown in Fig. 4. This capability allows image synthesis with a fine-tuned image contrast, adapted to a use case and contrast requirements. Varying TE significantly influences the muscle tissue's signal intensity, while varying TR mainly causes signal intensity differences within the joint [20]. The TE and TR estimate from the auxiliary classifier show that the contrast settings were correctly reconstructed.

**Table 4** Quantitative evaluation results of the image synthesis experiments

| Model | NMSE | PSNR | SSIM |
|---|---|---|---|
| 1 | 0.13 ± 0.13 | 22.91 ± 2.61 | 0.62 ± 0.12 |
| 2 | 0.12 ± 0.14 | 23.45 ± 2.7 | 0.63 ± 0.11 |
| 3 | 0.10 ± 0.11 | 24.15 ± 2.58 | 0.64 ± 0.12 |
| 4 | 0.10 ± 0.11 | <u>24.29 ± 2.59</u> | <u>0.66 ± 0.11</u> |
| 5 | <u>0.09 ± 0.10</u> | <u>24.48 ± 2.81</u> | <u>0.66 ± 0.11</u> |
| 6 | 0.16 ± 0.21 | 22.14 ± 2.57 | 0.58 ± 0.12 |

Statistically significant ($p \leq 0.05$, two-sample t-test) best values are underlined.

Moreover, our approach correctly reconstructs different anatomies and acquisition parameter values, as demonstrated in Fig. 5.

## Discussion

Sequence parameterizations vary considerably across different sites, scanners, and scans in clinical practice (compare Fig. 2 and [1, 21, 22]). Consequently, MR image synthesis approaches must be able to cope with this variability to be widely applicable. Therefore, we have adapted our model architecture and training process to fit these needs by injecting the acquisition parameters into the model. Acquisition parameter injection has improved the reconstruction results, and our method outperforms a benchmark pix2pix approach on fat-saturated image synthesis from non-fat-saturated MR images. Additionally, we can adapt MR image contrast retrospectively and continuously using acquisition parameter injection using adaptive instance normalization. Our approach is the first to provide this capability.

To the best of our knowledge, our work is the second to apply an image-to-image translation network to knee MR images for contrast synthesis [8]. Comparison to published approaches (compare Section "Literature Review") is not meaningful due to different datasets used and the training on different body regions. Training our work on other benchmark datasets, e.g., the BRATS dataset [23], was not possible as acquisition parameters are not provided in the dataset.

However, there are also limitations within this work to point out. A limitation of the cycle consistency loss is that it may hallucinate features in the generated images [24]. It also decreases the overall reconstruction performance, likely due to the imperfection of the auxiliary classifier. However, it also increases the generalization ability of the network. It enables the network to learn a broader range of contrast settings as a larger set of input-target label combinations can be used for training. The use of unpaired data will likely be obsolete given a more diverse dataset. The diagnostic accuracy must be evaluated, in particular when using cycle consistency/unpaired data training.

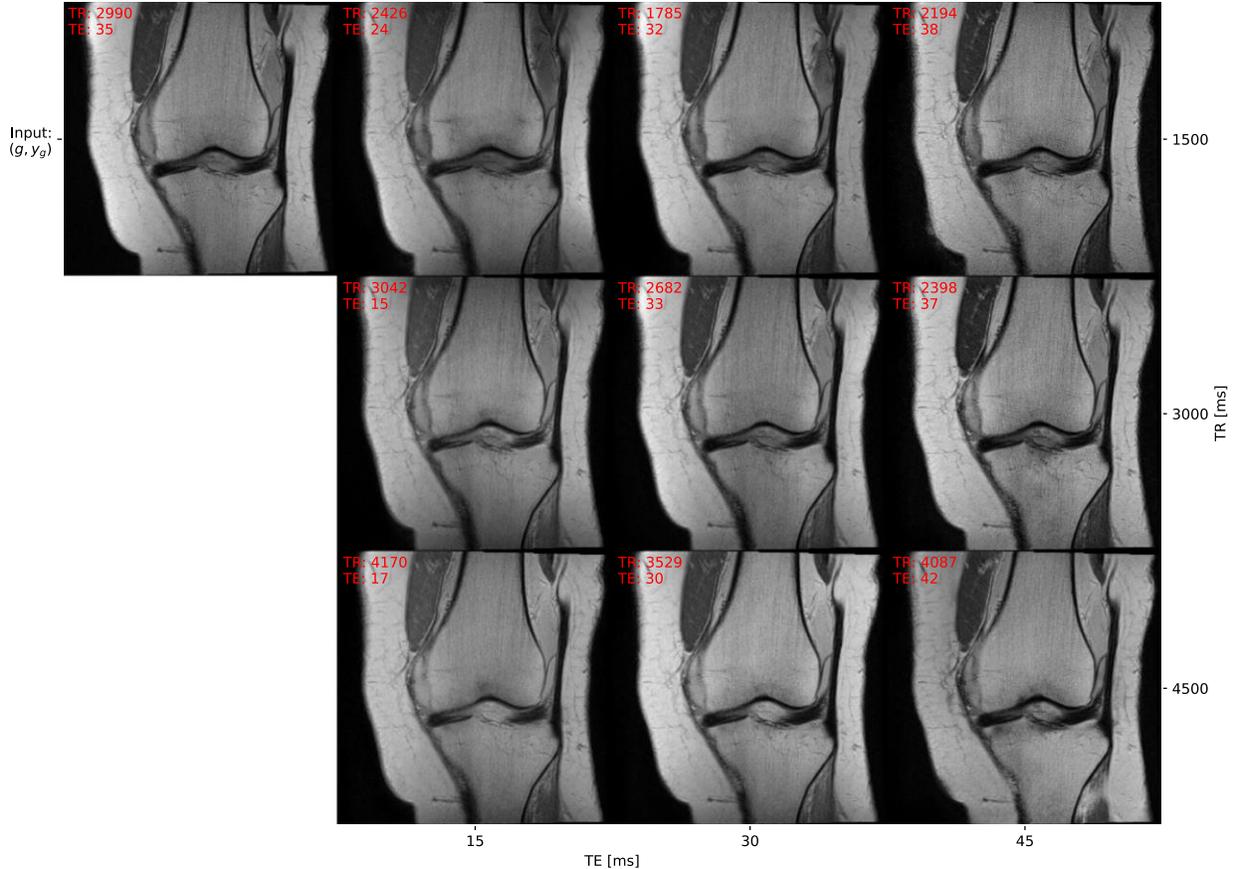

**Fig. 4** Example of contrast interpolation given a real non-fat-saturated MR image (top left). The images are annotated with the true acquisition parameters (TE, TR) for the input image, and with the acquisition parameters determined by the AC for the reconstructed examples.

We have also trained a model to translate fat-saturated images to non-fat-saturated images. Due to the lower SNR of fat-saturated MR images, results were significantly worse as the network must hallucinate more information and are therefore not suited for a true estimation of the target contrast. The dataset contains only PD-weighted non-fat-saturated and PD- and T2-weighted fat-saturated MR images based on the DICOM series description. Consequently, T1-weighted and T2-weighted MR images cannot be synthesized due to the lack of training data. Nonetheless, it is anticipated that our approach's capabilities can be easily extended given a more diverse (in terms of acquisition parameter value distribution) training dataset.

Furthermore, image pairs within the dataset are often not appropriately registered, and through-plane motion can be observed between paired image series. 2D rigid registration methods did not improve the training performance. However, efforts to incorporate 3D non-rigid image registration may increase data quality and thus model performance.

The GAN (model 6) receives feedback from the auxiliary classifier, a trained neural network, on how well the MR contrast was synthesized. Incorporating the AC allows us to train the GAN under a cycle consistency loss with random target labels. This is necessary as paired training data (especially pairs of non-fat-saturated MR images) are limited and not evenly distributed within the label space. However, while introducing unpaired training data and training under a cycle consistency loss utilizing an auxiliary classifier for the conditioning loss improves the capabilities of the model to synthesize MR images of arbitrary contrast, it has also limitations. CycleGANs are known to hallucinate features in medical image translations [24], and PSNR and SSIM values decrease w.r.t. only training on paired image data. Consequently, using a paired image dataset with a broader set of aquisiton parameter combinations is anticipated to significantly increase the capabilities of our approach to synthesize arbitrary MR contrasts and increase the reconstruction performance metrics.

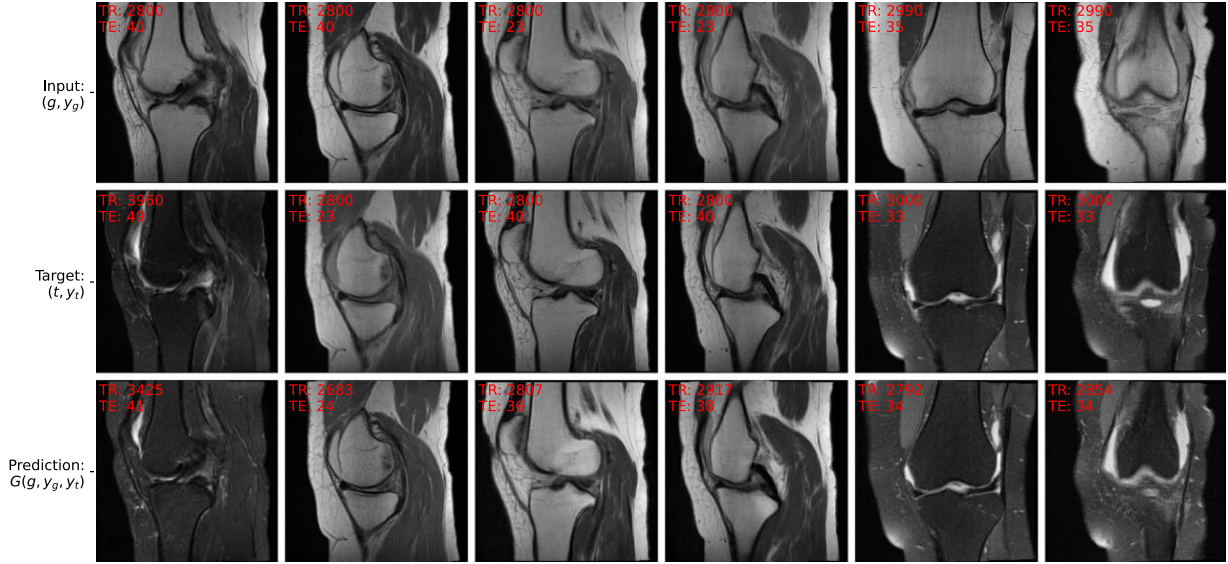

**Fig. 5** Multiple examples of image pairs and the reconstructed images using our proposed approach. All images belong to a single, reference test patient, randomly selected and ordered anatomically, by image series and slice number. The images are annotated as described for the previous figures. The model properly reconstructs a wide range of anatomical structures and views, with different sets of acquisition parameters.

Moreover, a general limitation of MR synthesis methods is that different MR sequences provide both redundant and unique information. Thus, the presented results are anticipated to improve significantly by using a multi-image-to-image translation approach.

Our work can be applied to several use cases. First, it is anticipated to be used as a method to replace missing or corrupted contrasts. This may shorten overall scan time and reduce the number of re-scans. Furthermore, it can be used as an advanced data augmentation tool to render different contrasts and increase the robustness of a trained AI application for contrast changes. It may also support the radiologists during protocoling by providing a preview of the contrast of a parameterized sequence. Additionally, this approach can also be useful for other image-to-image translation tasks within medical imaging, e.g., to enhance intermodality translation (MRI $\rightarrow$ CT) or 7T image synthesis from 3T MR images [25].

Our approach can be the basis for image-to-image tasks in the MRI domain as it is anticipated to be adaptable to any set of parameterized sequences. It is easily adaptable to additional inputs, different acquisition parameters, and applications.

In future work, we aim to improve the determination of the acquisition parameters, which will enhance contrast synthesis and enable us to evaluate our model more accurately. Moreover, the extension to 3D and multi-image-to-image contrast synthesis is anticipated to improve the performance as more data and information generally lead to an improved model. Furthermore, a reader study must be conducted to quantify the diagnostic value of the synthesized images and assess our approach's clinical significance for applications such as the synthesis of additional contrasts.

## Conclusion

In this paper, we proposed a novel approach to tackle MR image-to-image synthesis by guiding the learning process with the MR acquisition parameters that define the image tissue contrast. Our approach was evaluated on the task of synthesizing fat-saturated MR images from non-fat-saturated images, outperforming a pix2pix benchmark method, and demonstrated its capabilities for continuous contrast synthesis.

Our approach is easily extendible to incorporate more acquisition parameters, 3D image data, and multi-image-to-image translations that can further increase the reconstruction accuracy.

## Acknowledgment

This work is supported by the Bavarian Academic Forum (BayWISS)—Doctoral Consortium "Health Research",

funded by the Bavarian State Ministry of Science and the Arts.